\pdfoutput=1
\newcommand*{\ATLASLATEXPATH}{}
\documentclass[UKenglish,texlive=2017,cernpreprint,texmf]{\ATLASLATEXPATH atlasdoc}
\usepackage[subfigure,backend=biber]{\ATLASLATEXPATH atlaspackage}
\usepackage{\ATLASLATEXPATH atlasbiblatex}
\usepackage{\ATLASLATEXPATH atlascontribute}
\usepackage{\ATLASLATEXPATH atlasphysics}
\usepackage{a4wide}
\graphicspath{{/}{logos/}{figures/}}

\AtlasTitle{A measurement of the soft-drop jet mass in $pp$ collisions at $\sqrt{s}=13$~\TeV~with the ATLAS detector}
\author{The ATLAS Collaboration}
\AtlasJournalRef{Phys. Rev. Lett. 121 (2018) 092001}
\AtlasDOI{10.1103/PhysRevLett.121.092001}
\PreprintIdNumber{CERN-EP-2017-231}

\AtlasAbstract{%                                                                                                                       
Jet substructure observables have significantly extended the search program for physics beyond the Standard Model at the Large Hadron Collider.  The state-of-the-art tools have been motivated by theoretical calculations, but there has never been a direct comparison between data and calculations of jet substructure observables that are accurate beyond leading-logarithm approximation. Such observables are significant not only for probing the collinear regime of QCD that is largely unexplored at a hadron collider, but also for improving the understanding of jet substructure properties that are used in many studies at the Large Hadron Collider.  This Letter documents a measurement of the first jet substructure quantity at a hadron collider to be calculated at next-to-next-to-leading-logarithm accuracy.  The normalized, differential cross-section is measured as a function of $\log_{10}{\rho^2}$, where $\rho$ is the ratio of the soft-drop mass to the ungroomed jet transverse momentum.  This quantity is measured in dijet events from 32.9~fb$^{-1}$ of $\sqrt{s}=13$~\TeV\ proton--proton collisions recorded by the ATLAS detector.  The data are unfolded to correct for detector effects and compared to precise QCD calculations and leading-logarithm particle-level Monte Carlo simulations.
}

\hypersetup{pdftitle={ATLAS document},pdfauthor={The ATLAS Collaboration}}

\begin{document}

\maketitle

The dynamics of strong interactions, described by quantum chromodynamics (QCD), are responsible for most of the physical processes occurring in proton--proton ($pp$) scattering at the Large Hadron Collider (LHC).  The fundamental particles of QCD, quarks and gluons, cannot be observed directly and instead form collimated sprays of particles called \textit{jets} when produced at high energy.   The radiation pattern inside jets has been used extensively for identifying highly Lorentz boosted hadronically decaying massive particles~\cite{expsearch}. Many of these techniques were motivated by recent advances in analytical calculations of jet substructure~\cite{Larkoski:2017jix}.  However, prior to this work, there has never been a direct comparison between collision data and calculations beyond the leading-logarithm (LL) accuracy of Parton Shower (PS) Monte Carlo (MC)  programs~\cite{Buckley:2011ms}.   The comparisons presented here begin the field of precision jet substructure, wherein data and calculations in the collinear regime of QCD can be used to test the modeling of final state radiation and maybe even extract fundamental parameters of the SM such as the strong coupling constant or the top quark mass~\cite{Hoang:2017kmk}.  Such precision understanding will also be essential to maximize the quantitative sensitivity of the LHC and future colliders to physics beyond the Standard Model.

Of particular importance is the \textit{jet mass}, defined as the norm of the four-momentum sum of constituents inside a jet.  The jet mass is a key jet substructure observable and is the most powerful tool for identifying Lorentz boosted hadronically decaying massive particles.  Unlike Lorentz boosted bosons or top quarks, the mass of generic quark and gluon jets is set by the fragmentation of highly virtual partons~\cite{Salam:2009jx}.  A complete prediction for mass or other variables beyond LL has not been possible due to the presence of \textit{non-global logarithms} (NGLs)~\cite{Dasgupta:2001sh}: resummation terms associated with particles that radiate out of, and then radiate back into, a jet.  These terms are formally present at next-to-leading-logarithm (NLL) accuracy and have prevented full comparisons of observables beyond LL.  However, using insights from modern analytical methods, the authors of Ref.~\cite{Dasgupta:2013ihk} introduced a new procedure to systematically remove soft and wide-angle radiation from the jet (\textit{grooming}) that is formally insensitive to NGLs.  This procedure was extended in Ref.~\cite{Larkoski:2014wba} to form the soft-drop grooming algorithm.  The calculation of the masses of jets that have the soft-drop procedure applied is insensitive to NGLs.  The distribution of the soft-drop mass has now been calculated at both next-to-leading order (NLO) with NLL~\cite{Marzani:2017mva,Marzani:2017kqd} and leading order (LO) with next-to-next-to-leading-logarithm (NNLL) accuracy~\cite{Frye:2016aiz,Frye:2016okc}. These are the most precise calculations for jet substructure at a hadron collider.

The soft-drop procedure acts on the clustering history of a sequential recombination jet algorithm~\cite{Ellis:1993tq}.  In these algorithms, all inputs to jet-finding start as a proto-jet and are combined pairwise using a distance metric in $y$-$\phi$ space\footnote{ATLAS uses a right-handed coordinate system with its origin at the interaction point in the center of the detector. Cylindrical coordinates $(r,\phi)$ are used in the transverse plane, $\phi$ being the azimuthal angle around the beam axis. The pseudorapidity is defined in terms of the polar angle $\theta$ as $\eta=-\ln\tan(\theta/2)$.  The rapidity, differences of which are invariant under longitudinal boosts also for massive particles, is defined as $y = \frac{1}{2} \ln [(E+p_z)/(E-p_z)]$.}.  When the smallest distance is above some threshold $R$ (called the jet radius), the algorithm terminates and the remaining proto-jets are the final jets.  The clustering history is the sequence of pairwise combinations that lead to a particular jet.  Jets at the LHC experiments are usually clustered using the anti-$k_t$ algorithm~\cite{Cacciari:2008gp}, which has the benefit of producing regularly shaped jets in $y$-$\phi$ space.  Even though anti-$k_t$ jets are useful experimentally, their clustering history does not mimic the angular-ordered PS~\cite{Ellis:1991qj} used in the related $k_t$~\cite{Catani:1993hr,Ellis:1993tq} and Cambridge--Aachen~\cite{Dokshitzer:1997in,Wobisch:1998wt} (C/A) algorithms.  The soft-drop algorithm starts by re-clustering an anti-$k_t$ jet's constituents with the C/A algorithm.   Next, the clustering tree is traversed from the latest branch to the earliest and at each node the following criterion is applied to proto-jets $j_1$ and $j_2$:

\begin{align}
\label{eq:softdrop}
\frac{\text{min}(p_\text{T,$j_1$},p_\text{T,$j_2$})}{p_\text{T,$j_1$}+p_\text{T,$j_2$}}> z_\text{cut}\left(\frac{\Delta R_{12}}{R}\right)^\beta,
\end{align}

\noindent where $p_\text{T}$ is the momentum of a jet transverse to the beam pipe, $z_\text{cut}$ and $\beta$ are algorithm parameters, and $\Delta R_{12} = \sqrt{(\Delta y)^2+(\Delta \phi)^2}$ is the distance in $y$-$\phi$ between the proto-jets. The parameter $z_\text{cut}$ sets the scale of the energy removed by the algorithm; $\beta$ tunes the sensitivity of the algorithm to wide-angle radiation. If the soft-drop condition in Eq.~(\ref{eq:softdrop}) is not satisfied, then the branch with the smaller $p_\text{T}$ is removed.  The procedure is then iterated on the remaining branch.  If the condition is satisfied at any node, the algorithm terminates.  As $\beta$ increases, the fraction of branches where the condition is satisfied increases, reducing the amount of radiation removed from the jet.  In the limit $\beta\rightarrow \infty$, the original jet is untouched.  The mass of the resulting jet is referred to as the soft-drop jet mass, $m^\text{soft drop}$.

This Letter presents a measurement of the soft-drop jet mass using 32.9 fb$^{-1}$ of $\sqrt{s}=13$~\TeV~$pp$ data collected in 2016 by the ATLAS detector, and the first comparison to predictions of jet substructure that are formally more accurate than the LL PS approximation.

ATLAS is a particle detector designed to achieve nearly a full $4\pi$ coverage in solid angle~\cite{PERF-2007-01}.  The inner tracking detector (ID) is inside a 2~T magnetic field and is designed to measure charged-particle trajectories up to  $|\eta|=2.5$.  Surrounding the ID are electromagnetic and hadronic calorimeters, which use liquid argon and lead, copper, or tungsten absorber for the electromagnetic and forward ($|\eta| > 1.7$) hadronic detectors, and scintillator-tile active material with steel absorber for the central ($|\eta| < 1.7$) hadronic calorimeter.

For this study, jets are clustered using the anti-$k_t$ jet algorithm with radius parameter $R=0.8$ implemented in FastJet~\cite{Cacciari:2011ma}. The inputs are topological calorimeter-cell clusters calibrated using the local cluster weighting algorithm~\cite{Aad:2016upy}. In order to improve the rapidity resolution, cluster four-vectors are corrected to point toward the reconstructed primary collision vertex~\cite{PERF-2015-01}. An overall  jet energy calibration, derived for $R=0.8$ jets, accounts for residual detector effects as well as contributions from pileup (i.e., simultaneous additional $pp$ collisions) in order to make the reconstructed jet energy unbiased (up through `Absolute MC-based calibration' in Ref.~\cite{Aaboud:2017jcu}).  Jets are required to have $|\eta| < 1.5$ so that their calorimeter-cell clusters are within the coverage of the ID. 

Events were selected online using a two-level trigger system~\cite{Aaboud:2016leb} that is hardware-based at the first level and software-based for the second level. In this analysis, the full-luminosity jet trigger with the lowest \pt threshold is nearly 100\% efficient for jets with $\pt > 600$~\GeV. Events are required to have a minimum of two jets, at least one of which has $p_\text{T}> 600$~\GeV.  In addition, a dijet topology is imposed by requiring that the leading two $p_\text{T}$-ordered jets satisfy $p_\text{T,1}/p_\text{T,2}<1.5$: as the leading two jets are required to have similar $\pt$, this removes events with additional energetic jets.

The soft-drop algorithm is then run on the leading two jets in the selected events.  Both of these jets are used for the measurement. Three different values of $\beta \in \{0, 1, 2\}$ are considered. The value of $z_\text{cut}$ is fixed at $0.1$ so that $\log(z_\text{cut})$ resummation is negligible~\cite{Marzani:2017mva}. The dimensionless mass $\rho = m^\text{soft drop} / \pt^\text{ungroomed}$ is the observable of interest: as the soft-drop mass is correlated with \pt, $\rho$ is a dimensionless quantity that only weakly depends on \pt. For each $\beta$ value, $\log_{10}(\rho^2)$ is constructed from the jet's mass after the soft drop algorithm and its \pt before (referred to as $\pt^\text{ungroomed}$).  The ungroomed jet \pt is used because the groomed version is collinear unsafe when $\beta=0$~\cite{Marzani:2017mva}.  The full $\log_{10}(\rho^2)$ distribution is studied, but the focus is on the \textit{resummation region} ($-3.7<\log_{10}(\rho^2)<-1.7$), where resummation dominates over non-perturbative or fixed-order parts of the recent precision calculations; studying the distribution in log-scale allows this region to be studied more closely.

After the event selection, the data are unfolded to correct for detector effects.  MC simulations are used to perform the unfolding and for comparisons with the corrected data.   The unfolding procedure corrects detector-level\footnote{Detector level refers to the measured outputs of the detector; particle level refers to the particles that interact with the detector.} observables to particle-level. The particle-level selection is defined to be as close as possible to the detector-level selection in order to minimize the size of simulation-based corrections when unfolding.  Particle-level jets are clustered from simulated particles with a mean lifetime $\tau > 30 $~ps excluding muons and neutrinos.  These jets are built using the same algorithm as for detector-level jets, and particle-level events must pass the same dijet requirement. The experimental resolution of the $\log_{10}(\rho^2)$ distribution depends on the jet $\pt$, so the $\log_{10}(\rho^2)$ and $\pt$ distributions are simultaneously unfolded. After correcting for the acceptance of the event selection, the full two-dimensional distribution is unfolded using an iterative Bayesian (IB) technique~\cite{DAgostini:1994zf} with four iterations as implemented in the RooUnfold framework~\cite{Adye:2011gm}.   The acceptance corrections are largely independent of $\log_{10}(\rho^2)$, with a small effect below $-3$ due to the $\rho\neq 0$ requirement. 

Several MC simulations are used to unfold and compare to the data. Dijet events were generated at LO using \PYTHIA~\cite{Sjostrand:2007gs} 8.186, with the $2 \rightarrow 2$ matrix element (ME) convolved with the NNPDF2.3LO parton distribution function (PDF) set~\cite{Carrazza:2013axa}, and using the A14~\cite{ATL-PHYS-PUB-2014-021} set of tuned PS and underlying-event model parameters. Additional radiation beyond the ME was simulated in \PYTHIA 8 using the LL approximation for the \pt-ordered PS~\cite{Corke:2010zj}. To provide several comparisons to data, additional dijet samples were simulated using different generators. \SHERPA 2.1.1~\cite{Gleisberg:2008ta} generates events using multi-leg $2\rightarrow 3$ matrix elements, which are matched to the PS following the CKKW prescription~\cite{Catani:2001cc}. These \SHERPA events were simulated using the CT10 LO PDF set~\cite{Lai:2010vv} and the default \SHERPA event tune. \textsc{Herwig++} 2.7.1~\cite{Bahr:2008pv,Corcella:2000bw} events were generated with the $2\rightarrow 2$ matrix element, convolved with the CTEQ6L1 PDF set~\cite{Pumplin:2002vw} and configured with the UE-EE-5 tune~\cite{Gieseke:2012ft}.  Both \SHERPA and \textsc{Herwig++} use angular ordering in the PS and a cluster model for hadronization~\cite{Webber:1983if}.  All MC samples use \PYTHIA 8 minimum bias events (MSTW2008LO PDF set~\cite{mstw} and A2 tune~\cite{ATL-PHYS-PUB-2012-003}) to simulate pileup. They were processed using the full ATLAS detector simulation~\cite{Aad:2010ah} based on \GEANT4~\cite{Agostinelli:2002hh}.

\begin{figure}[h!]
\begin{center}
\includegraphics[width=0.49\textwidth]{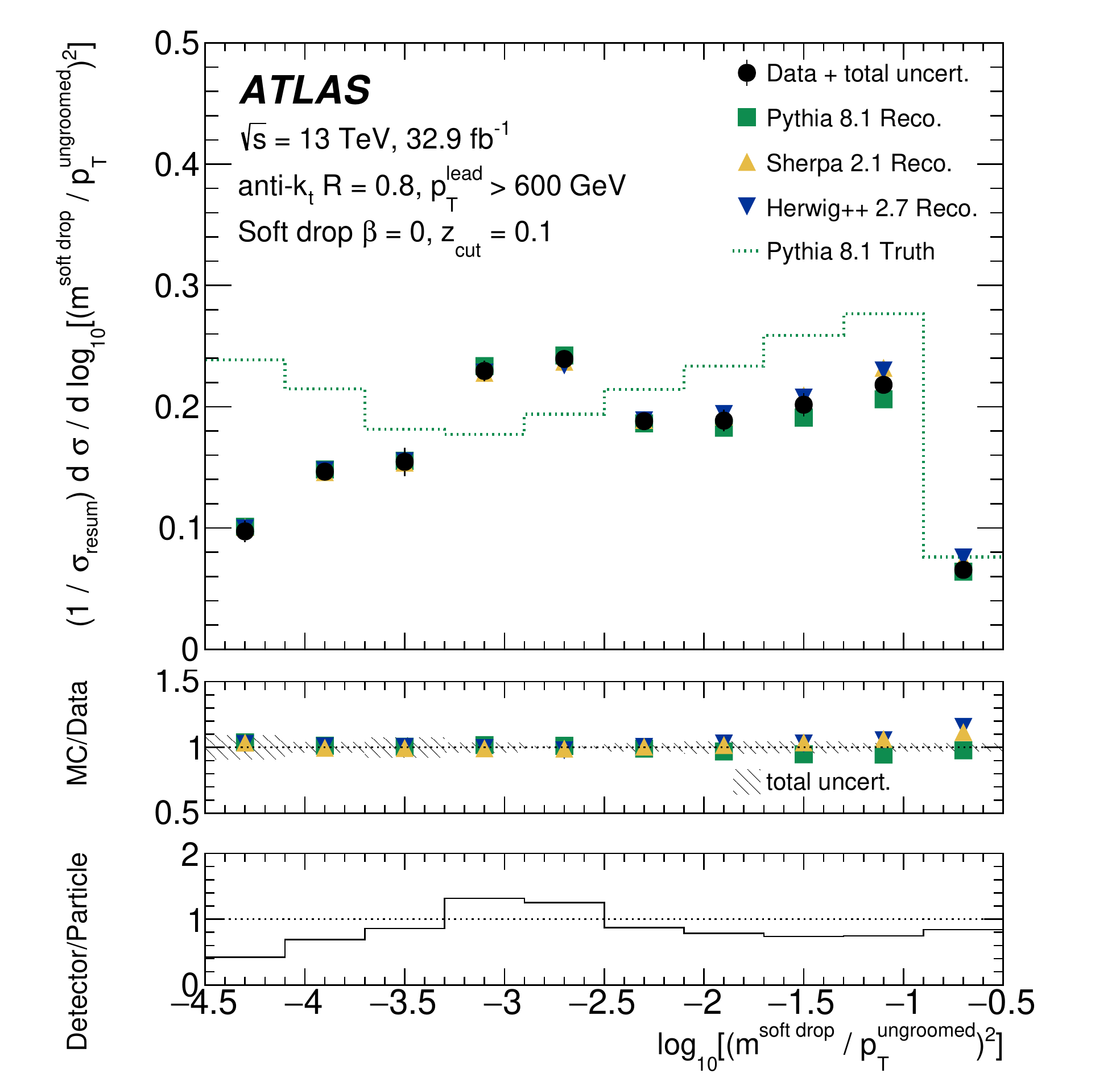}
\includegraphics[width=0.49\textwidth]{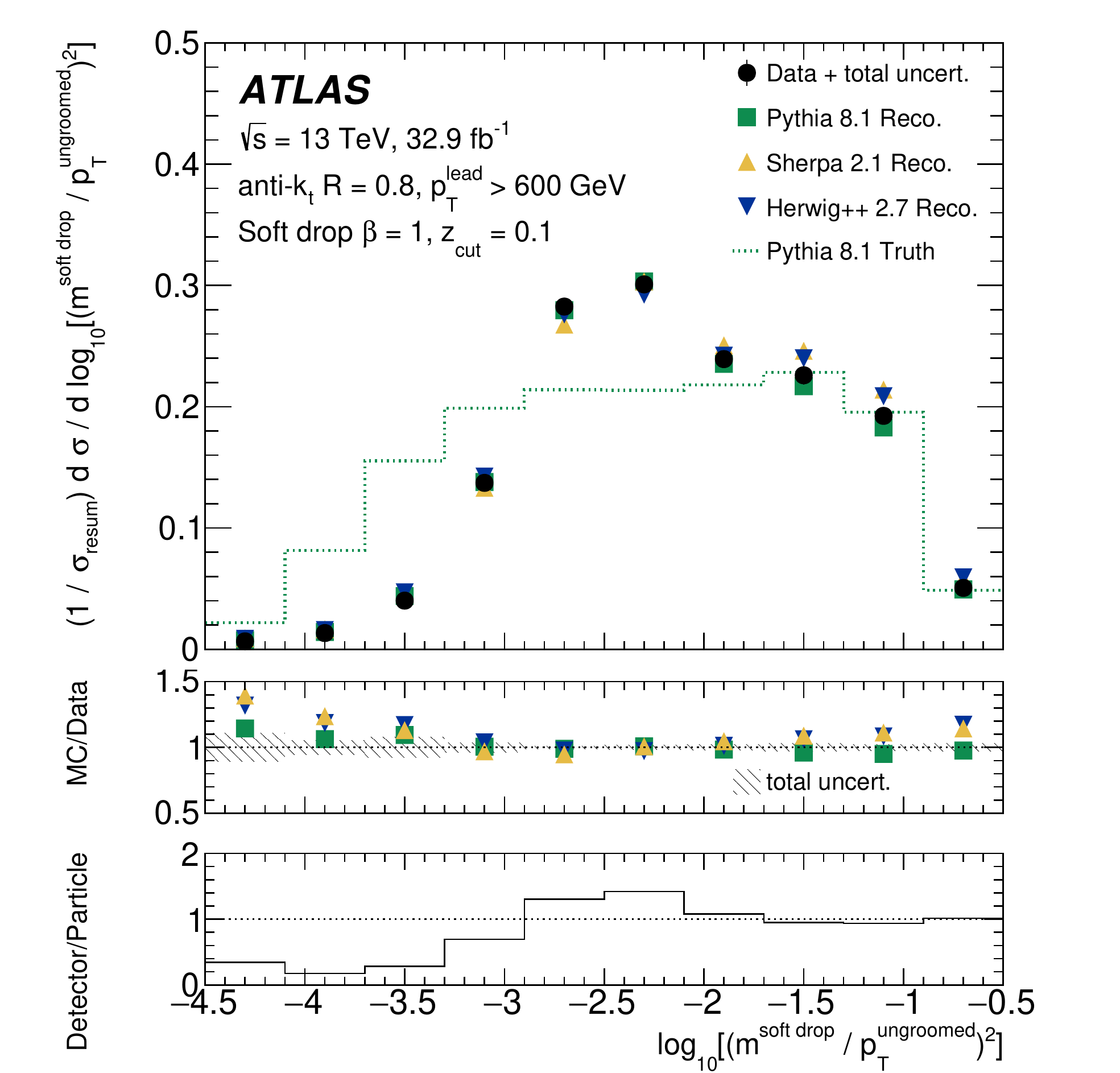}\\
\includegraphics[width=0.49\textwidth]{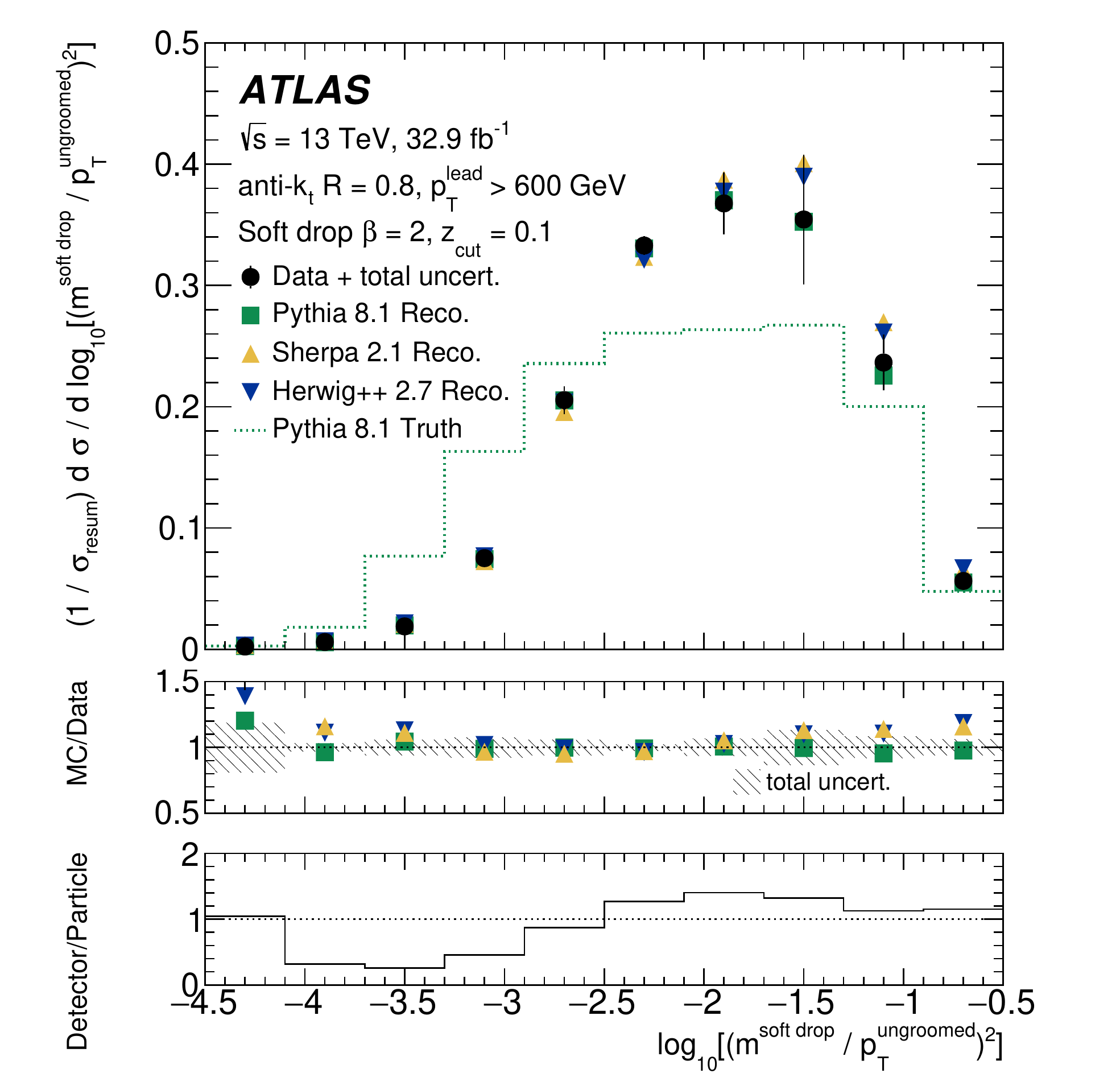}
\end{center}
\caption{Distributions of $\log_{10}(\rho^2)$ in data compared to reconstructed detector-level (Reco.) \textsc{Pythia}, \textsc{Sherpa}, and \textsc{Herwig++}, and particle-level (Truth) \textsc{Pythia} simulations for $\beta=0$ (left), $\beta=1$ (right), and $\beta=2$ (bottom).  The ratio of the three detector-level MC predictions to the data is shown in the middle panel, and the size of the detector$\rightarrow$ particle-level corrections for \textsc{Pythia} is shown as the ratio in the bottom panel. The error bars on the data points and in the first ratio include the experimental systematic uncertainties in the cluster energy, angular resolution, and efficiency. The distributions are normalized to the integrated cross-section, $\sigma_{\text{resum}}$, measured in the resummation region, $-3.7 < \log_{10}(\rho^2) < -1.7$.  }
\label{fig:unfolding:theorycompare}
\end{figure}

 Figure~\ref{fig:unfolding:theorycompare} shows the uncorrected data compared with detector-level simulation for \textsc{Pythia}, \textsc{Sherpa}, and \textsc{Herwig++} as well as particle-level simulation for \textsc{Pythia}.  There are substantial migrations between the detector- and particle-level distributions, which cause large off-diagonal terms in the unfolding matrix especially at low values of $\log_{10}(\rho^2)$.

Various systematic uncertainties impact the soft-drop mass distribution. The sources of uncertainty can be classified into two categories: experimental and theoretical modeling.  Experimental uncertainties are due to limitations in the accuracy of the modeling of calorimeter-cell cluster energies and positions as well as their reconstruction efficiency, and are evaluated as follows.  Isolated calorimeter-cell clusters are matched to tracks; the mean and standard deviation of the energy-to-momentum ratio ($E/p$) is used for the cluster energy scale and resolution uncertainties, and the standard deviation of the relative position is used for the cluster angular resolution.  In the track-momentum range $30$~\GeV $<p<350$~\GeV, $E/p$ is augmented with information from testbeam studies~\cite{Abat:2010zza}.  For $|\eta| > 0.6$ in that $p$ range or for $p > 350$~\GeV~(and any $|\eta|$), a flat 10\% uncertainty is estimated for both the energy scale and resolution, motivated by earlier studies~\cite{Aaboud:2016hwh}.  The reconstruction efficiency is studied using the fraction of tracks without a matched calorimeter-cell cluster.  A series of validation studies are performed to ensure that these uncertainties are valid also for non-isolated clusters.  Jets clustered from tracks are geometrically matched to calorimeter jets and the ratio of their $p_\text{T}$ and mass is sensitive to the jet energy scale (JES) and jet mass scale.  Furthermore, the decomposition method~\cite{ATLAS-CONF-2010-052,Aad:2012vm,Aaboud:2016hwh} is used to propagate the cluster-based uncertainties to an effective JES, which agrees well with the observed in-situ shift for $R=0.4$ ungroomed jets~\cite{Aaboud:2017jcu}.  Finally, the jet mass scale and resolution are tested using the observed $W$ mass peak in $t\bar{t}$ events.  The same event selection and level of agreement is observed as in Ref.~\cite{ATLAS-CONF-2017-063}.  These additional studies confirm that the cluster-based uncertainties are valid for $\log_{10}(\rho^2)$.

One of the dominant uncertainties is due to the theoretical modeling of jet fragmentation (QCD Modeling). In particular, as dijet simulation is used to unfold the data, the results of the analysis are sensitive to the choice of MC generator used for this procedure. The \textsc{Pythia} generator is used for the nominal sample, and comparisons are made with \textsc{Sherpa} and \textsc{Herwig++}. The \textsc{Sherpa} and \textsc{Herwig++} generators give compatible results, so only the variation with \textsc{Sherpa} is used as a systematic uncertainty. The impact of this uncertainty is assessed by unfolding the data with the alternative response matrix. In addition to directly varying the model used to derive the response matrix, a data-driven nonclosure technique is used to estimate the potential bias from a given choice of prior and the number of iterations in the IB method~\cite{Malaescu:2009dm}. The inverse of the response matrix is applied to the particle-level spectrum, which is reweighted until the folded spectrum agrees with data. This modified detector-level distribution is unfolded with the nominal response matrix and the difference between this and the reweighted particle-level spectrum is taken as an uncertainty. Finally, the sensitivity of the unfolding procedure to pile-up is assessed by reweighting events to vary the distribution of the number of interactions in the MC simulation by 10\%: the impact on the measurement is small. This is expected, since the soft-drop algorithm is designed to remove the soft, wide-angle radiation that pileup contributes.

\begin{figure}
\begin{center}
\includegraphics[width=0.8\textwidth]{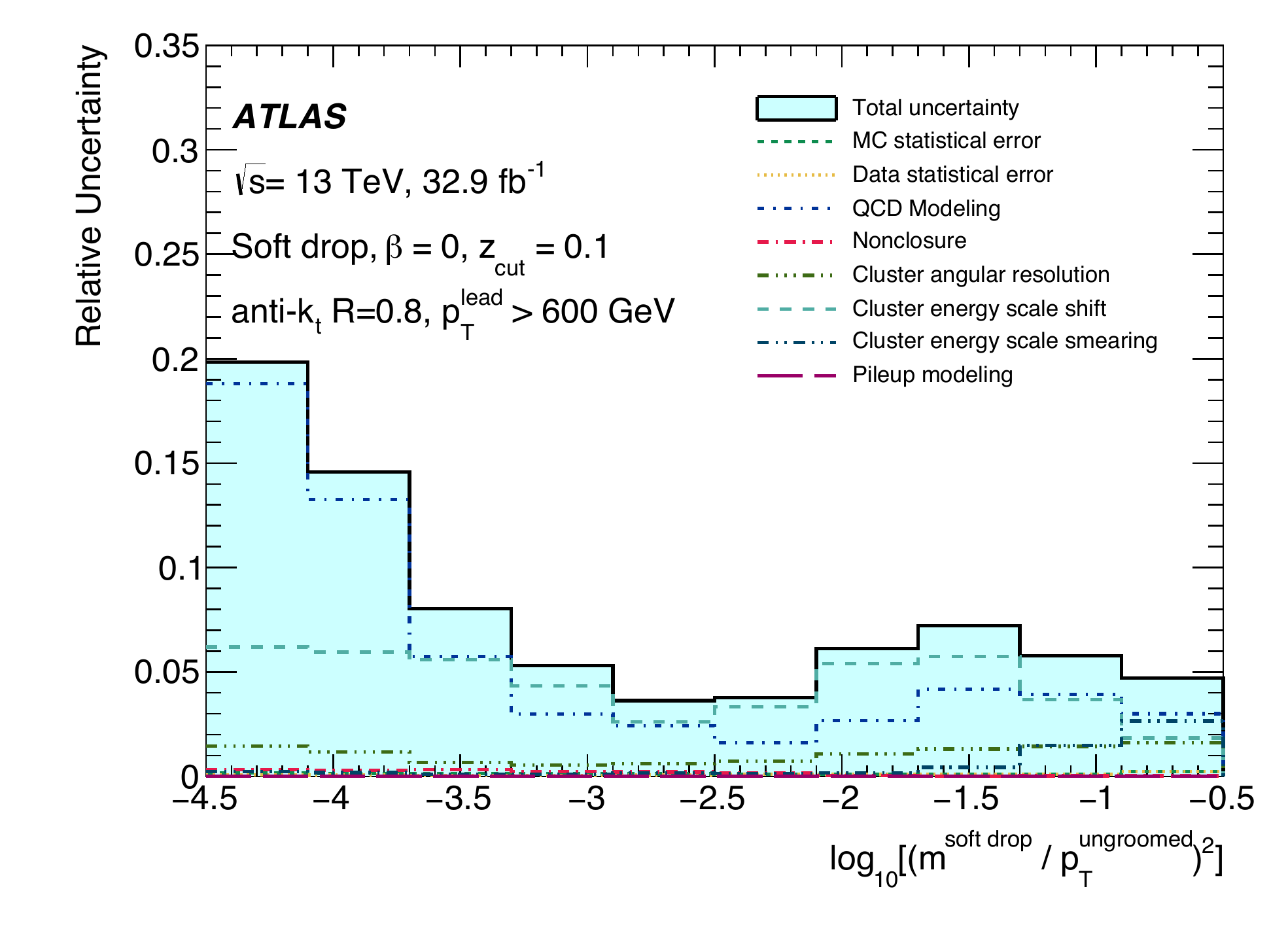}
\end{center}
\caption{The breakdown of systematic and statistical uncertainties as a function of $\log_{10}(\rho^2)$ for $\beta=0$.}
\label{results:uncertainties0}
\end{figure}

The uncertainties are dominated by QCD modeling and the cluster energy scale.  The former are largest ($\lesssim 20\%$) at low $\log_{10}(\rho^2)$ where non-perturbative effects introduce a sensitivity to the $\log_{10}(\rho^2)$ distribution prior, and are $\lesssim 10\%$ for the rest of the distribution.  Cluster energy uncertainties are large ($\lesssim 5\%$) at low $\log_{10}(\rho^2)$ where the cluster multiplicity is low and also at high $\log_{10}(\rho^2)$ where the energy of the hard prongs, rather than their opening angle, dominates the mass resolution.   Other sources of uncertainty are typically below $5\%$ across the entire distribution.  A summary of the relative sizes of the various systematic uncertainties for $\beta = 0$ is shown in Fig.~\ref{results:uncertainties0}.  The relative sizes of the different sources of systematic uncertainty are similar for $\beta = 1$ and $\beta=2$, except that the large uncertainty at low $\log_{10}(\rho^2)$ values spans a larger range.

\begin{figure}[h!]
\centering
\includegraphics[width=0.5\textwidth]{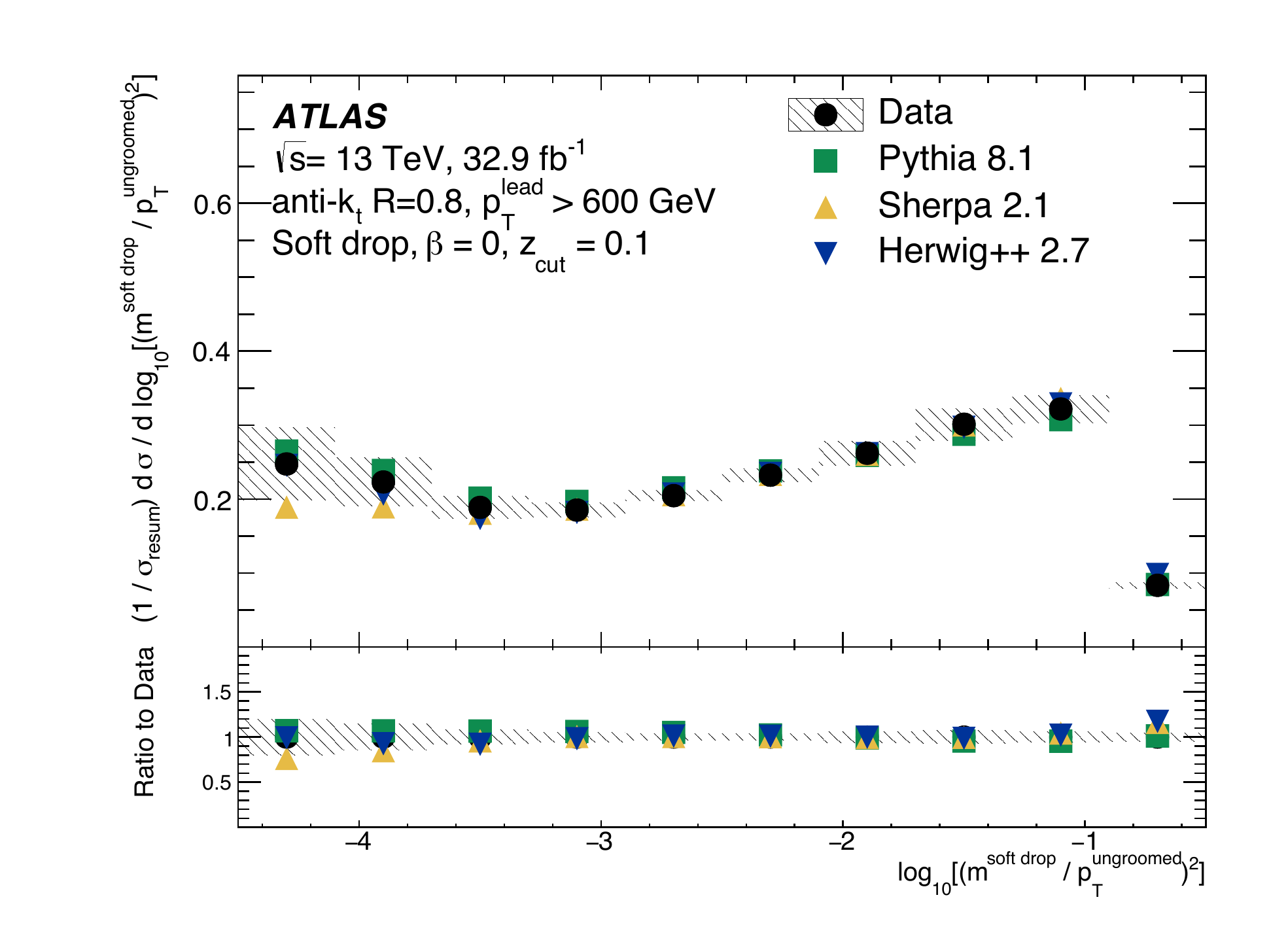}\includegraphics[width=0.5\textwidth]{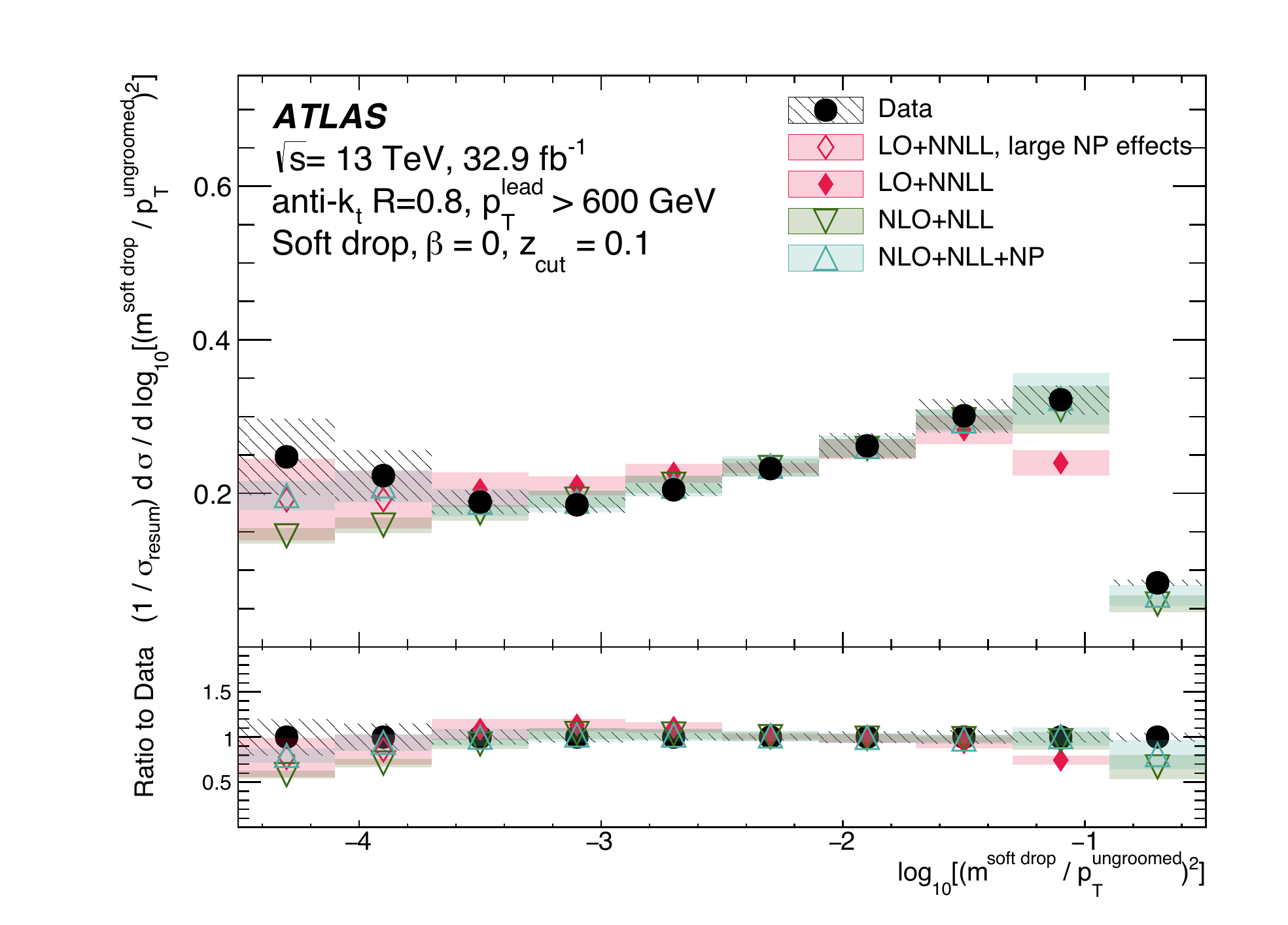}
\includegraphics[width=0.5\textwidth]{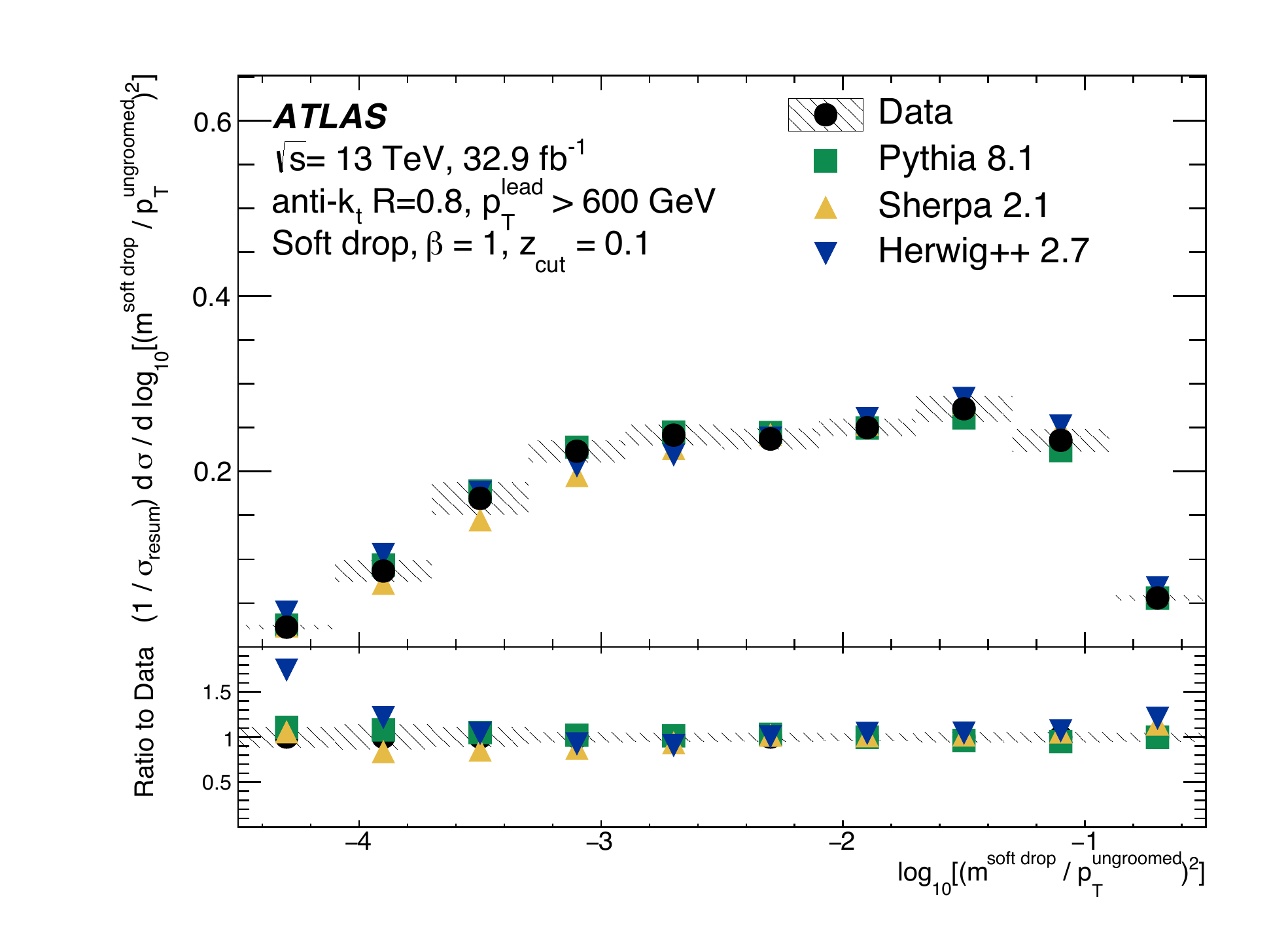}\includegraphics[width=0.5\textwidth]{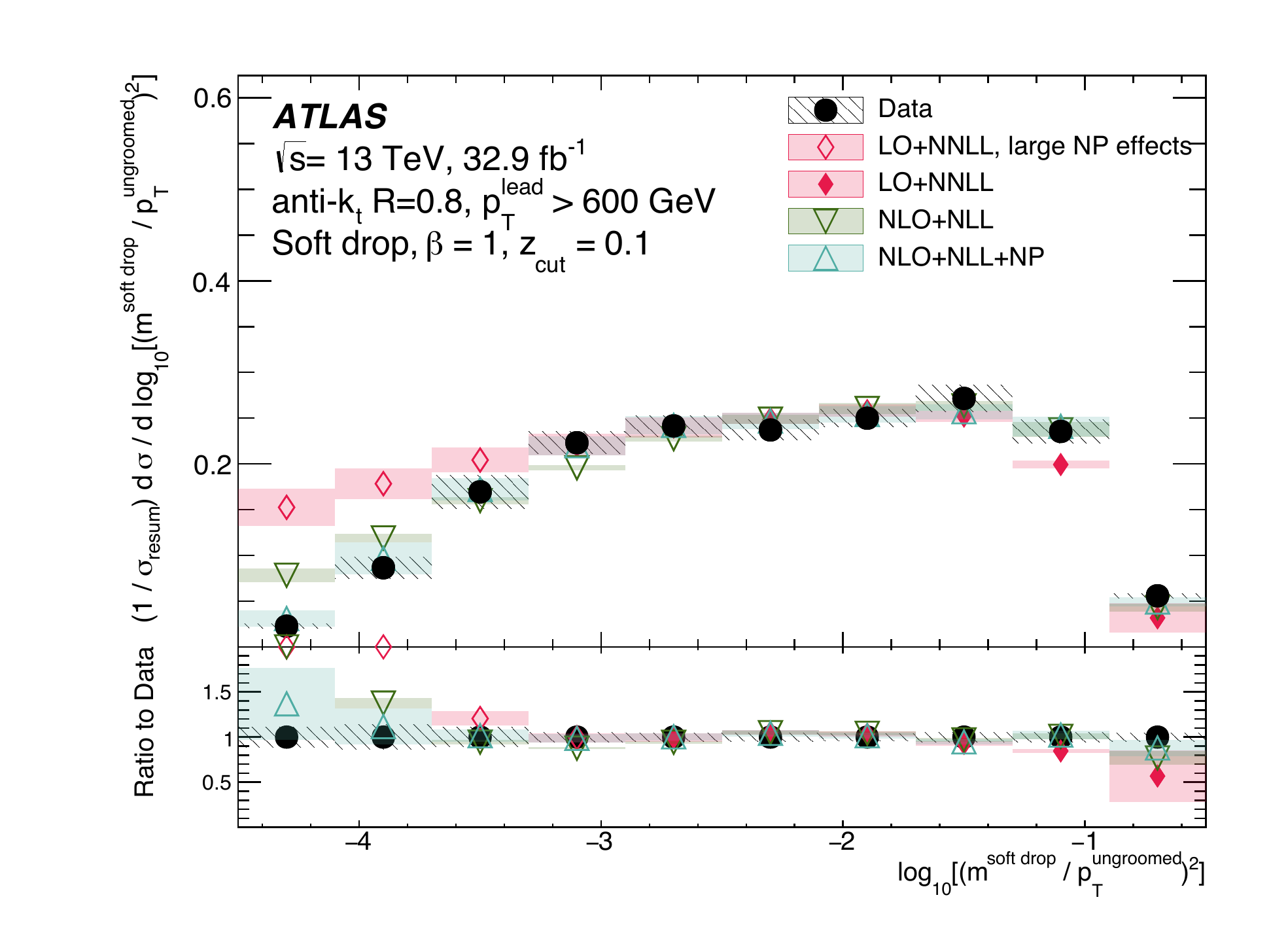}
\includegraphics[width=0.5\textwidth]{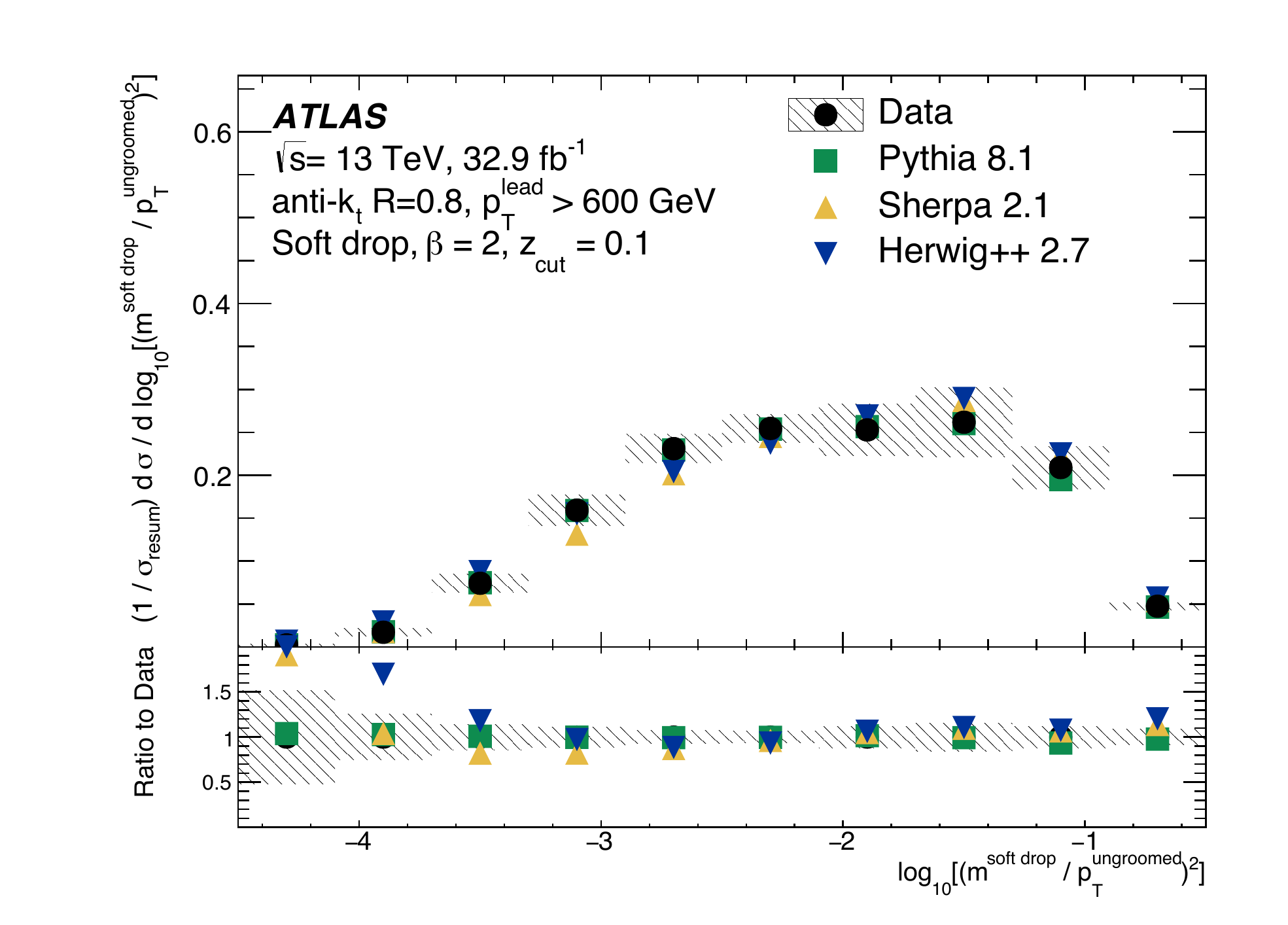}\includegraphics[width=0.5\textwidth]{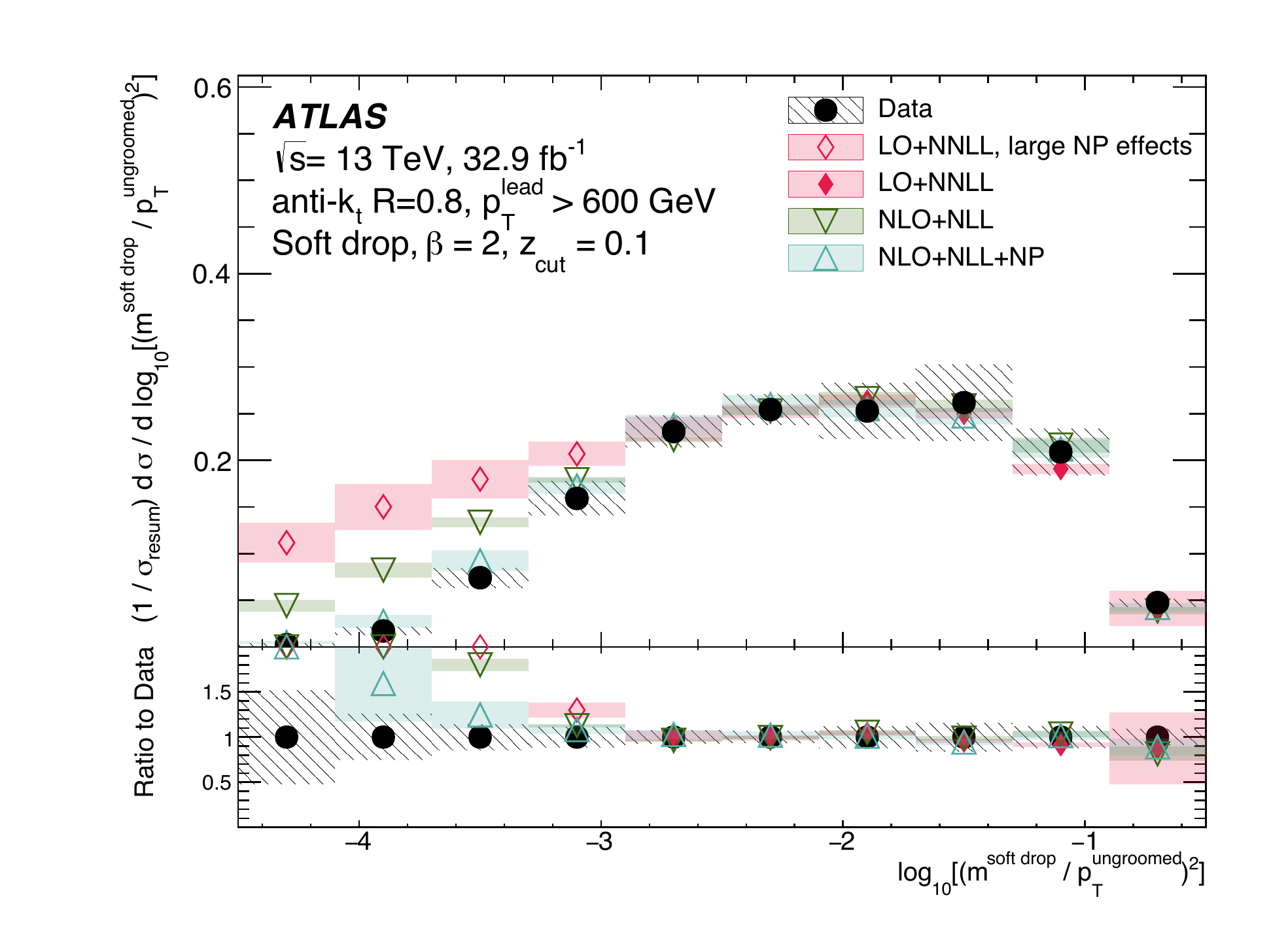}
\caption{The unfolded $\log_{10}(\rho^2)$ distribution for anti-$k_t$ $R=0.8$ jets with $p_\text{T}^\text{lead}$ > 600 GeV, after the soft drop algorithm is applied for $\beta \in \{0, 1, 2\}$, in data compared to \textsc{Pythia}, \textsc{Sherpa}, and \textsc{Herwig++} particle-level (left), and NLO+NLL(+NP) \cite{Marzani:2017mva} and LO+NNLL \cite{Frye:2016okc,Frye:2016aiz} theory predictions (right). The LO+NNLL calculation does not have non-perturbative (NP) corrections; the region where these are expected to be large is shown in a open marker (but no correction is applied), while regions where they are expected to be small are shown with a filled marker.  All uncertainties described in the text are shown on the data; the uncertainties from the calculations are shown on each one. The distributions are normalized to the integrated cross section, $\sigma_{\text{resum}}$, measured in the resummation region, $-3.7 < \log_{10}(\rho^2) < -1.7$.  The NLO+NLL+NP cross-section in this resummation regime is 0.14,  0.19, and 0.21 nb for $\beta=0,1,2$, respectively~\cite{Marzani:2017mva}.}
\label{results:unfold:integral}
\end{figure}

The unfolded data are shown in Fig.~\ref{results:unfold:integral}. They are compared to the predictions of the \textsc{Pythia}, \textsc{Sherpa}, and \textsc{Herwig++} generators, as well as the NLO+NLL prediction from Ref.~\cite{Marzani:2017mva,Marzani:2017kqd} and the LO+NNLL prediction from Refs.~\cite{Frye:2016aiz,Frye:2016okc}.   The (N)NLL calculations use NLOJet++~\cite{Catani:1996vz,Nagy:2003tz} (MG5\_aMC~\cite{Alwall:2014hca}) with the CT14nlo~\cite{Dulat:2015mca} (MSTW2008LO) PDF set for matrix element calculations.  The distributions are normalized to the integrated cross-section, $\sigma_{\text{resum}}$, measured in the resummation region, $-3.7 < \log_{10}(\rho^2) < -1.7$. The uncertainties due to the analytical calculation come from independently varying each of the renormalization, factorization, and resummation scales by factors of $2$ and $1/2$.   The NLO+NLL calculation is also given with non-perturbative (NP) corrections based on the average of various MC models with NP effects turned on and off; the envelope of predictions is added as an uncertainty~\cite{Marzani:2017mva}.  The LO+NNLL predictions do not contain NP effects, but the open makers in Fig.~\ref{results:unfold:integral} indicate where NP are expected to be large (`large NP effects').

The MC predictions and the analytical calculations are expected to be accurate in different regions of $\log_{10}(\rho^2)$~\cite{Marzani:2017mva,Frye:2016okc,Frye:2016aiz}. In general, non-perturbative effects are large for $\log_{10}(\rho^2) < -3.7$ (where small-angle or soft gluon emissions dominate) and small for $-3.7 < \log_{10}(\rho^2) < -1.7$ where resummation dominates.  Fixed, higher-order corrections are expected to be important for $\log_{10}(\rho^2) > -1.7$, where large-angle gluon emission can play an important role. This implies that the region $-3.7 < \log_{10}(\rho^2) < -1.7$ (the resummation region) should have the most reliable predictions for both the MC generators and the LO+NNLL analytical calculation, while the NLO+NLL calculation should also be accurate for $\log_{10}(\rho^2) > -1.7$. For all values of $\beta$, the measured and predicted shapes agree well in the resummation region, and the data and NLO+NLL prediction continue to agree well at higher values of $\log_{10}(\rho^2)$. At more negative values of $\log_{10}(\rho^2)$, non-perturbative effects lead to distinctly different predictions between the MC generators and the calculations without NP corrections; the data fall below the predictions for all $\beta$ values.  Interestingly, the NNLL calculation is not everywhere a better model of the data than the NLL calculation in the resummation regime and NP effects can also be comparable to the higher order resummation corrections in this regime.  Therefore, improved precision for the future will require will require a careful comparative analysis of the different perturbative calculations as well as a deeper and possibly analytic understanding of NP effects.

As $\beta$ increases, the fraction of radiation removed by soft-drop grooming decreases and the impact of non-perturbative effects grows larger~\cite{Frye:2016okc,Frye:2016aiz}, so the range over which the analytical calculations are accurate also decreases. The degree of agreement between data and all the calculations for $\log_{10}(\rho^2) < -3$ does substantially worsen for $\beta \in \{1, 2\}$, especially when NP corrections are not included.  Agreement between the data and the MC generators remains generally within uncertainties for all values of $\beta$.  Digitized versions of the results, along with versions binned in jet \pt can be found at Ref.~\cite{hepdata}.

\FloatBarrier 

In summary, a measurement of the soft-drop jet mass is reported. The measurement provides a comparison of the internal properties of jets between 32.9 fb$^{-1}$ of 13~\TeV~$pp$ collision data collected by the ATLAS detector at the LHC and precision QCD calculations accurate beyond leading logarithm. Where the calculations are well defined perturbatively, they agree well with the data; in regions where non-perturbative effects are expected to be significant, the calculations disagree with the data and the predictions from MC simulation are better able to reproduce the data. The dijet cross section is presented as a normalized fiducial dijet differential cross section as a function of the $\log_{10}(\rho^2)$ for each jet, allowing the results to be used to constrain future calculations and MC generator predictions.

%-------------------------------------------------------------------------------
\section*{Acknowledgments}
%-------------------------------------------------------------------------------

% Acknowledgements for papers with collision data
% Version 14-Feb-2018

% Standard acknowledgements start here                                                                                                                                                    
%----------------------------------------------                                                                                                                                                
We thank CERN for the very successful operation of the LHC, as well as the
support staff from our institutions without whom ATLAS could not be
operated efficiently.

We acknowledge the support of ANPCyT, Argentina; YerPhI, Armenia; ARC, Australia; BMWFW and FWF, Austria; ANAS, Azerbaijan; SSTC, Belarus; CNPq and FAPESP, Brazil; NSERC, NRC and CFI, Canada; CERN; CONICYT, Chile; CAS, MOST and NSFC, China; COLCIENCIAS, Colombia; MSMT CR, MPO CR and VSC CR, Czech Republic; DNRF and DNSRC, Denmark; IN2P3-CNRS, CEA-DRF/IRFU, France; SRNSFG, Georgia; BMBF, HGF, and MPG, Germany; GSRT, Greece; RGC, Hong Kong SAR, China; ISF, I-CORE and Benoziyo Center, Israel; INFN, Italy; MEXT and JSPS, Japan; CNRST, Morocco; NWO, Netherlands; RCN, Norway; MNiSW and NCN, Poland; FCT, Portugal; MNE/IFA, Romania; MES of Russia and NRC KI, Russian Federation; JINR; MESTD, Serbia; MSSR, Slovakia; ARRS and MIZ\v{S}, Slovenia; DST/NRF, South Africa; MINECO, Spain; SRC and Wallenberg Foundation, Sweden; SERI, SNSF and Cantons of Bern and Geneva, Switzerland; MOST, Taiwan; TAEK, Turkey; STFC, United Kingdom; DOE and NSF, United States of America. In addition, individual groups and members have received support from BCKDF, the Canada Council, CANARIE, CRC, Compute Canada, FQRNT, and the Ontario Innovation Trust, Canada; EPLANET, ERC, ERDF, FP7, Horizon 2020 and Marie Sk{\l}odowska-Curie Actions, European Union; Investissements d'Avenir Labex and Idex, ANR, R{\'e}gion Auvergne and Fondation Partager le Savoir, France; DFG and AvH Foundation, Germany; Herakleitos, Thales and Aristeia programmes co-financed by EU-ESF and the Greek NSRF; BSF, GIF and Minerva, Israel; BRF, Norway; CERCA Programme Generalitat de Catalunya, Generalitat Valenciana, Spain; the Royal Society and Leverhulme Trust, United Kingdom.

The crucial computing support from all WLCG partners is acknowledged gratefully, in particular from CERN, the ATLAS Tier-1 facilities at TRIUMF (Canada), NDGF (Denmark, Norway, Sweden), CC-IN2P3 (France), KIT/GridKA (Germany), INFN-CNAF (Italy), NL-T1 (Netherlands), PIC (Spain), ASGC (Taiwan), RAL (UK) and BNL (USA), the Tier-2 facilities worldwide and large non-WLCG resource providers. Major contributors of computing resources are listed in Ref.~\cite{ATL-GEN-PUB-2016-002}.
%----------------------------------------------

\printbibliography

\newpage % ATLAS Collaboration author list
% Data extracted on 02-Sep-2017 for paper reference STDM-2017-04
\begin{flushleft}
{\Large The ATLAS Collaboration}

\bigskip

M.~Aaboud$^\textrm{\scriptsize 137d}$,
G.~Aad$^\textrm{\scriptsize 88}$,
B.~Abbott$^\textrm{\scriptsize 115}$,
O.~Abdinov$^\textrm{\scriptsize 12}$$^{,*}$,
B.~Abeloos$^\textrm{\scriptsize 119}$,
S.H.~Abidi$^\textrm{\scriptsize 161}$,
O.S.~AbouZeid$^\textrm{\scriptsize 139}$,
N.L.~Abraham$^\textrm{\scriptsize 151}$,
H.~Abramowicz$^\textrm{\scriptsize 155}$,
H.~Abreu$^\textrm{\scriptsize 154}$
% ...
\bigskip
\\
% ...
$^{12}$ Institute of Physics, Azerbaijan Academy of Sciences, Baku, Azerbaijan\\
% ...
$^{88}$ CPPM, Aix-Marseille Universit{\'e} and CNRS/IN2P3, Marseille, France\\
% ...
$^{115}$ Homer L. Dodge Department of Physics and Astronomy, University of Oklahoma, Norman OK, United States of America\\
% ...
$^{119}$ LAL, Univ. Paris-Sud, CNRS/IN2P3, Universit{\'e} Paris-Saclay, Orsay, France\\
% ...
$^{137}$ $^{(a)}$ Facult{\'e} des Sciences Ain Chock, R{\'e}seau Universitaire de Physique des Hautes Energies - Universit{\'e} Hassan II, Casablanca; $^{(b)}$ Centre National de l'Energie des Sciences Techniques Nucleaires, Rabat; $^{(c)}$ Facult{\'e} des Sciences Semlalia, Universit{\'e} Cadi Ayyad, LPHEA-Marrakech; $^{(d)}$ Facult{\'e} des Sciences, Universit{\'e} Mohamed Premier and LPTPM, Oujda; $^{(e)}$ Facult{\'e} des sciences, Universit{\'e} Mohammed V, Rabat, Morocco\\
% ...
$^{139}$ Santa Cruz Institute for Particle Physics, University of California Santa Cruz, Santa Cruz CA, United States of America\\
% ...
$^{151}$ Department of Physics and Astronomy, University of Sussex, Brighton, United Kingdom\\
% ...
$^{154}$ Department of Physics, Technion: Israel Institute of Technology, Haifa, Israel\\
$^{155}$ Raymond and Beverly Sackler School of Physics and Astronomy, Tel Aviv University, Tel Aviv, Israel\\
% ...
$^{161}$ Department of Physics, University of Toronto, Toronto ON, Canada\\
% ...
$^{a}$ Also at Department of Physics, King's College London, London, United Kingdom\\
$^{b}$ Also at Institute of Physics, Azerbaijan Academy of Sciences, Baku, Azerbaijan\\
$^{c}$ Also at Novosibirsk State University, Novosibirsk, Russia\\
$^{d}$ Also at TRIUMF, Vancouver BC, Canada\\
$^{e}$ Also at Department of Physics {\&} Astronomy, University of Louisville, Louisville, KY, United States of America\\
% ...
$^{*}$ Deceased
\end{flushleft}

\end{document}